\documentclass[prl, superscriptaddress, twocolumn]{revtex4-1}
\usepackage{amsfonts}
\usepackage{amssymb}
\usepackage[dvips]{graphicx}
\usepackage{amsmath}
\usepackage{textcomp}

\newcommand{\IdealState}{\varrho_{\text{id}}}
\newcommand{\ExpState}{\varrho_{\text{exp}}}
\newcommand{\PrivateState}{\varrho_{\text{priv}}}

\newcommand{\ket}[2][]{{|#2\rangle_{#1}}}
\newcommand{\bra}[2][]{{}_{#1}\langle #2|}

\newcommand{\proj}[2][]{\ket{#2}_{#1}\bra{#2}}

\newcommand{\Tr}{\textrm{Tr}}

\begin{document}
\title{Experimental extraction of secure correlations from a noisy private state}
\author{K.~Dobek}
\affiliation{Institute of Physics, Nicolaus Copernicus University, ul.~Grudziadzka 5/7, 87-100 Toru\'{n}, Poland}
\affiliation{Faculty of Physics, Adam Mickiewicz University, ul.~Umultowska 85, 61-614 Pozna{\'n}, Poland}
\author{M.~Karpi{\'n}ski}
\affiliation{Faculty of Physics, University of Warsaw, ul.~Ho\.{z}a 69, 00-681 Warsaw, Poland}
\author{R.~Demkowicz-Dobrza{\'n}ski}
\affiliation{Faculty of Physics, University of Warsaw, ul.~Ho\.{z}a 69, 00-681 Warsaw, Poland}
\author{K.~Banaszek}
\affiliation{Institute of Physics, Nicolaus Copernicus University, ul.~Grudziadzka 5/7, 87-100 Toru\'{n}, Poland}
\affiliation{Faculty of Physics, University of Warsaw, ul.~Ho\.{z}a 69, 00-681 Warsaw, Poland}
\author{P.~Horodecki}
\affiliation{Faculty of Applied Physics and Mathematics, Technical University of Gda{\'n}sk, ul. Narutowicza 11/12, 80-952 Gda{\'n}sk, Poland}


\date{\today}
\begin{abstract}
We report experimental generation of a noisy entangled four-photon state that exhibits a separation between the secure key contents and distillable entanglement, a hallmark feature of the recently established quantum theory of private states. The privacy analysis, based on the full tomographic reconstruction of the prepared state, is utilized in a proof-of-principle key generation. The inferiority of distillation-based strategies to extract the key is demonstrated by an implementation of an entanglement distillation protocol for the produced state.
\end{abstract}
\maketitle

Quantum entanglement can guarantee secure communication as demonstrated by Ekert's protocol \cite{EkertPRL1991}
for quantum key distribution \cite{GisinRibordyRMP2002} 
(QKD), where the random key obtained from a maximally entangled state is known exclusively to legitimate users. A natural way to realise QKD using imperfect noisy entanglement is to attempt its distillation into the maximal form using local operations and classical communication \cite{BennettBrassardPRL1996}. 
This strategy however may reduce the attainable key length or even preclude its generation altogether, which follows from the recently developed theory of private quantum states \cite{HorodeckiHorodeckiPRL2005}.
The secure key can be extracted in general at higher rates than that implied by distillable entanglement, and even from certain classes of bound entangled states.

In this Letter we report experimental generation and utilization of a noisy entangled
four-photon state that exhibits the separation between secure key contents and distillable
entanglement. We perform a full tomographic reconstruction
of the produced state using the maximum-likelihood \cite{BanaszekDArianoPRA1999} 
and Bayesian reconstruction methods \cite{BuzekDerkaAnP1998, AudenaertScheelNJP2009},
which allows us to obtain credible estimates for the
quantities of interest despite their nonlinear character and high sensitivity to statistical noise and experimental imperfections. We present a proof-of-principle extraction of a secure key and implement an entanglement distillation protocol verified to perform suboptimally.

The original example of extracting privacy from quantum entanglement is Ekert's QKD protocol, in which two communicating parties---Alice and Bob---need a sequence of bipartite systems prepared in a maximally entangled two-qubit state such as $\ket{\phi_+} = \frac{1}{\sqrt{2}}\bigl(\ket{00} + \ket{11}\bigr)$. Local projections performed by Alice and Bob in the computational basis $\ket{0}, \ket{1}$ yield perfectly correlated random key bits. The security is checked by measuring the qubits in superposition bases to test coherence between the components $\ket{00}$ and $\ket{11}$. If the state used for QKD is indeed pure, the monogamy of entanglement \cite{Coffman2000} prevents an eavesdropper Eve from learning measurement outcomes obtained by legitimate users. Of course, a state $\ket{\phi_-} = \frac{1}{\sqrt{2}}\bigl(\ket{00} - \ket{11}\bigr)$ would be equally suitable for key generation. But an equiprobable statistical mixture of $\ket{\phi_+}$ and $\ket{\phi_-}$ ensures no security. This is because it can be viewed as a partial trace $\frac{1}{2}\bigl(\proj[AB]{\phi_-} + \proj[AB]{\phi_+} \bigr) = \Tr_{E} \bigl(\proj[ABE]{\Phi}\bigr)$ of a tripartite state
\begin{equation}
\label{Eq:PhiABE}
\ket[ABE]{\Phi} = \frac{1}{\sqrt{2}} \bigl( \ket[AB]{00}\otimes \ket[E]{0} + \ket[AB]{11}\otimes \ket[E]{1} \bigr)
\end{equation}
involving a qubit $E$ in possession of Eve, who can gain complete information about the results of Alice's and Bob's measurements in the computational basis without introducing any disturbance.

Suppose now that in addition to qubits $A$ and $B$, Alice and Bob possess also qubits $A'$ and $B'$ prepared jointly in a statistical mixture of
$\ket[AB]{\phi_-} \otimes \ket[A'B']{00}$
and
$\ket[AB]{\phi_+} \otimes \ket[A'B']{11}$.
Obviously, a local measurement of $A'$ or $B'$ in the computational basis reveals whether the qubits $A$ and $B$ have been prepared in $\ket{\phi_+}$ or $\ket{\phi_-}$. This enables key generation and entanglement distillation with equal rates. An intriguing case is the privacy of a mixed four-qubit state \cite{HorodeckiHorodeckiPRL2005}:
\begin{equation}
\label{Eq:AA'BB'phi+-}
\PrivateState\! = {\textstyle \frac{1}{4}} \proj[AB]{\phi_-} \otimes \varrho_{-}^{A'B'}
+  {\textstyle \frac{3}{4}} \proj[AB]{\phi_+} \otimes \varrho_{+}^{A'B'},
\end{equation}
where $\varrho_-\!\!  = \proj{\psi_-}$, $\varrho_+\!\! = \frac{1}{3}(\openone - \proj{\psi_-})$, and
we denote $\ket{\psi_{\pm}} = \frac{1}{\sqrt{2}} \bigl( \ket{01} \pm \ket{10} \bigr)$. Unlike the preceding example, the two operators $\varrho_\pm^{A'B'}$ cannot be discriminated unambiguously by Alice and Bob using local operations and classical communication, which lowers the value of distillable entanglement $E_D$ \cite{BennettDiVincenzoPRA1996}. This can be seen from an upper bound
\begin{equation}
E_D \le {\cal L} = \log_2 \Tr |\PrivateState^\Gamma| = \log_2 3 -1 \approx 0.585,
\end{equation}
where ${\cal L}$ is the log-negativity \cite{VidalWernerPRA2002} calculated for the partial transposition
$^\Gamma$ with respect to the partition $AA':BB'$. In contrast, the theory of private states \cite{HorodeckiHorodeckiPRL2005}---of which $\PrivateState$ is an example---shows that results of projecting qubits $A$ and $B$ in the computational basis cannot be learnt by Eve, thus providing one bit of a secure key. This leads to a gap between the key rate and $E_D$, implying general sub-optimality of distillation strategies.

\begin{figure}[t]
\includegraphics[width=8.5cm]{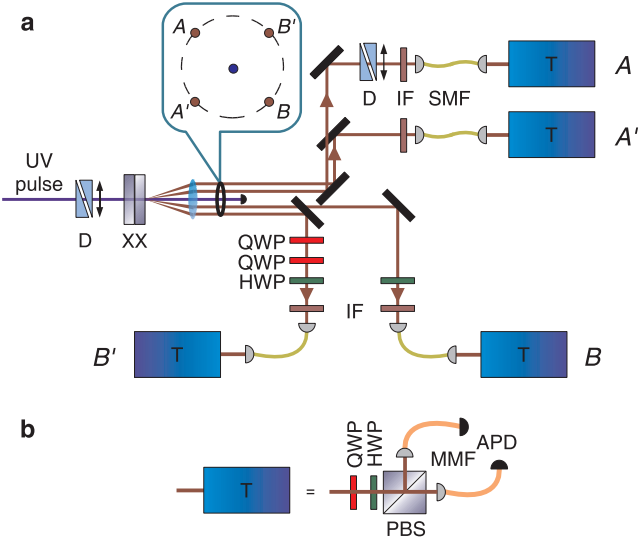}
\caption{Experimental setup. (a) Preparation of noisy private states. Two maximally entangled
photon pairs are generated in two nonlinear crystals XX, collected from four directions $AA'BB'$ shown in the inset, and subjected to polarization transformations implemented with quarter-wave plates QWP and half-wave plates HWP. D, Soleil-Babinet compensators; IF, interference filters; SMF, single-mode fibers. (b) Polarization analyzers. PBS, polarizing beam splitter; MMF, multi-mode fibers; APD, avalanche photodiodes.}
\label{fig:setup}
\end{figure}

In order to demonstrate experimentally this hallmark feature of private states
we generated a noisy entangled four-photon states
using a setup shown in Fig.~\ref{fig:setup}.
Its heart were two $1~\mathrm{mm}$ long type-I down conversion beta-barium borate crystals with optical axes aligned in perpendicular planes,
following the arrangement introduced by Kwiat {\em et al.} \cite{Kwiat1999}. The crystals
were pumped using  Ti:sapphire oscillator (Coherent Chameleon Ultra) emitting a $78~\mathrm{MHz}$ train of $180~\mathrm{fs}$ pulses frequency doubled in a $1~\mathrm{mm}$ long lithium triborate crystal to give a $390~\mathrm{nm}$ wavelength pump of an average power of $200~\mathrm{mW}$,
and focused to a $70$~\textmu m diameter waist.
The axial symmetry of type-I down-conversion implies that photons emerging along any two opposite ends of the emission cone will be maximally entangled. That way one can collect multiple photon pairs, as shown in the inset of Fig.~\ref{fig:setup}(a), and obtain a four-photon state $\ket{\phi_+}_{AB} \otimes \ket{\phi_+}_{A'B'}$ with $\ket{0}$ and $\ket{1}$ corresponding to horizontal and vertical polarizations.
Collimated photons after transmission through $10~\mathrm{nm}$ full-width-at-half-maximum bandwidth interference filters were coupled into single-mode fibers wound on manual polarization controllers.
Phase relations between two-photon probability amplitudes were controlled by two Soleil-Babinet compensators D placed in the path of the pump beam and photons $A$.

Photons $B$ were sent through a half-wave plate whose two selected orientations introduced a transformation $\sigma_x = \ket{0}\bra{1} + \ket{1}\bra{0}$
or $\sigma_z = \proj{0} - \proj{1}$. The set of two quarter-waveplates and a half-wave plate placed in the path of photons $B'$ realized one of four operations $\openone$, $\sigma_x$, $\sigma_z$, or $\sigma_y = i \sigma_x \sigma_z$. Applying combinations $\sigma_z^B \otimes \sigma_y^{B'}$, $\sigma_x^B \otimes \openone^{B'}$, $\sigma_x^B \otimes \sigma_x^{B'}$, and $\sigma_x^B \otimes \sigma_z^{B'}$ randomly with equal probabilities produced ideally the state
\begin{equation}
\label{Eq:AA'BB'phi-psi+}
\IdealState = {\textstyle \frac{1}{4}} \proj[AB]{\phi_-} \otimes \varrho_{-}^{A'B'}
+  {\textstyle \frac{3}{4}} \proj[AB]{\psi_+} \otimes \varrho_{+}^{A'B'},
\end{equation}
equivalent up to a local unitary to $\PrivateState$. The secure key can be obtained by measuring qubits $A$ and $B$ in the eigenbasis of $\sigma_y$ given by
$\ket{\bar{v}} = \frac{1}{\sqrt{2}}\bigl(\ket{0} + i (-1)^v \ket{1}\bigr)$, $v=0,1$.

The photons were detected using free space polarization analyzers constructed from a quarter-wave plate, a half-wave plate and a Wollaston polarizer with two output ports coupled into multimode fibers, connected to single photon counting modules SPCM (Perkin-Elmer SPCM-AQRH), as shown in Fig.~\ref{fig:setup}(b). Detection efficiencies within each polarization analyzer, determined from an independent macroscopic measurement, were equalized in the postprocessing by binomial resampling. Electric signals from SPCMs were registered with a field programmable gate array (FPGA) circuit using a coincidence window of $6~\mathrm{ns}$. Typical count rates were $10^5~\mathrm{s}^{-1}$ for single counts, $6 \times 10^3~\mathrm{s}^{-1}$ for two-photon 
and $2~\mathrm{s}^{-1}$ for fourfold coincidences.

\begin{figure}
\includegraphics[width=8.5cm]{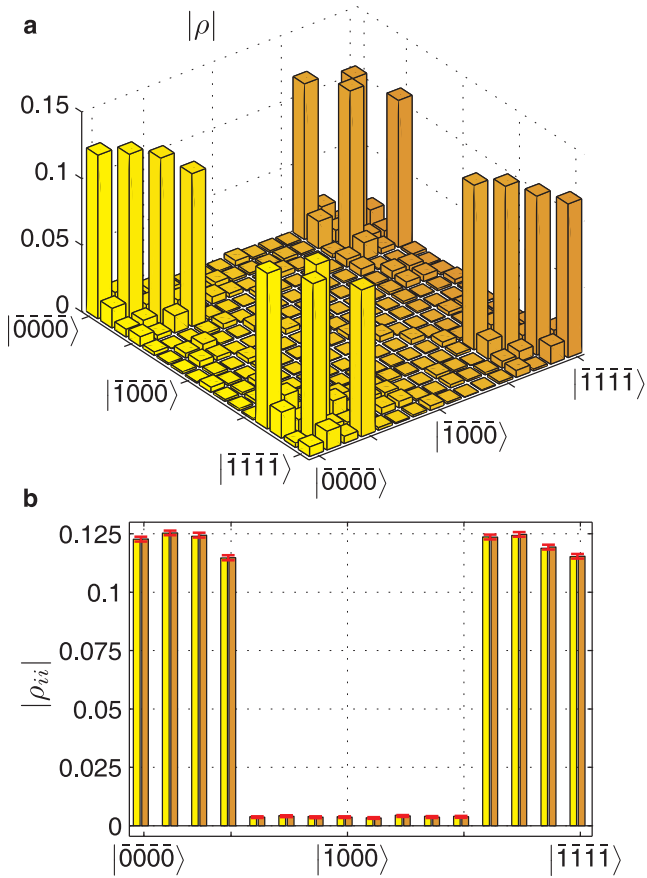}
\caption{Reconstructed private state. (a) absolute values
of density matrix elements in the $\sigma_y$ basis reconstructed using KF method. (b) diagonal KF values (orange, with error bars) compared with the ML results (yellow).}
\label{fig:rho}
\end{figure}

Assuming that only
four-photon events are available to Alice and Bob, we reconstructed a density matrix of a private state
  and performed a proof-of-principle secure key generation.
A complete measurement consisted of a sequence of
$33637$ intervals, each 10~s long. Before a single interval, settings of individual polarization analyzers were selected randomly and independently on Alice's and Bob's side to project polarization in the eigenbasis of $\sigma_x$, $\sigma_y$, or $\sigma_z$.
The density matrix of the generated state was reconstructed from fourfold coincidences using two independent techniques: the Kalman filter (KF) method \cite{AudenaertScheelNJP2009} based on gaussian approximation and Bayesian inference which provides an {\em a posteriori} probability distribution on the set of density matrices, and the maximum-likelihood (ML) method with physical constraints \cite{BanaszekDArianoPRA1999}.
In the KF approach the resulting {\em a posteriori} distribution served to generate a sample of $10^4$ physical density matrices with the help of the slice-sampling technique \cite{Mackay2003}. This sample was used to calculate mean values and standard deviations of individual elements of the density matrix depicted in Fig.~\ref{fig:rho}, as well as the information-theoretic quantities reported in Eq.~(\ref{Eq:ExpSeparation}).
Uncertainties of ML estimates were obtained by generating 2000 reconstructions using perturbed experimental data as an input.
The uncertainties calculated account for both the Poissonian photon counting noise and  $0.25^{\circ}$
 uncertainty of the waveplate orientation in polarization analyzers. Calculation of the KF a posteriori distribution took $20$~s on
a standard PC, a significant advantage compared with $20$~min
for the ML method. A more time consuming stage, however, was
generation of statistical samples of physical density matrices,
which took $2$~s per matrix using the KF distribution and
required repetition each time of the full reconstruction in the
ML case.

 Fig.~\ref{fig:rho} depicts the state $\ExpState$ obtained using the KF method.
The fidelity ${\cal F} = \Tr (\sqrt{\sqrt{\IdealState}\ExpState \sqrt{\IdealState}} )$
of this state is ${\cal F}_{\text{KF}} = 0.9724(7)$, and the ML value ${\cal F}_{\text{ML}} = 0.9715(7)$
lies within the confidence interval.
 The figure shows that the qubits $A$ and $B$ are indeed strongly correlated in the basis $\ket{\bar{0}}, \ket{\bar{1}}$.
To characterize the privacy of these correlations, we consider a purification $\ket[AA'BB'E]{\Psi}$ of the complete system $AA'BB'E$ in the worst-case scenario when Eve controls all environmental degrees of freedom $E$. Thus $\ExpState = \Tr_{E}\bigl(\proj[AA'BB'E]{\Psi}\bigr)$, which
generalizes Eq.~(\ref{Eq:PhiABE}).
After Alice projects the qubit $A$ onto a state $\ket{a}$, the state of Bob's qubit reduces to
$\varrho^{(a)}_B = \frac{1}{p_a}\Tr_{A'B'E} \bigl( {}_{A}\langle a \proj[AA'BB'E]{\Psi} a \rangle_{A} \bigr)$,
while Eve is in possession of a system in a state
$\varrho^{(a)}_E = \frac{1}{p_a} \Tr_{A'BB'} \bigl( {}_{A}\langle a \proj[AA'BB'E]{\Psi} a \rangle_{A} \bigr)$, where
$p_a = \Tr_{A'BB'E} \bigl( {}_{A}\langle a \proj[AA'BB'E]{\Psi} a \rangle_{A} \bigr)$ is the probability of obtaining the projection onto $\ket{a}$ by Alice. An attempt to gain information about Alice's outcome by either Bob or Eve can be viewed as a classical to quantum communication channel $A\rightarrow B$ or $A\rightarrow E$ \cite{DevetakWinterPRL2004}. In such a scenario---denoted as cqq---Alice and Bob can establish a secret key at a rate at least
\begin{equation}
\label{Eq:Keycqq}
{\cal X}^{\text{cqq}}  = \chi_{B} - \chi_{E},
\end{equation}
where $\chi_{B(E)}$ is the Holevo quantity \cite{HolevoPII1973} for the respective channel $A\rightarrow B(E)$, defined as:
\begin{equation}
\chi_{B(E)}=S\left(\sum_{a} p_a \rho^{(a)}_{B(E)}\right)- \sum_{a} p_a S\left(\rho^{(a)}_{B(E)}\right),
\end{equation}
$S(\cdot)$ denotes the von Neumann entropy, and the summations are carried out over an orthonormal basis
of states $\ket{a}$, in our case $\ket{\bar{0}}$ and $\ket{\bar{1}}$.

Based on measured data, the Bayesian {\em a posteriori} distribution for density matrices yields the following estimates for the
attainable key rate and the log-negativity
\begin{equation}
\label{Eq:ExpSeparation}
{\cal X}^{\text{cqq}}_{\text{KF}} =0.690(7),
\quad
{\cal L}_{\text{KF}}  = 0.581(4).
\end{equation}
These results show a clear separation, exceeding ten standard deviations, between distillable entanglement and the key rate, exposing a fundamental feature of general private states. The ML method yields consistent results ${\cal X}^{\text{cqq}}_{\text{ML}} = 0.704(7)$ and ${\cal L}_{\text{ML}} = 0.578(4)$.
The slightly higher value of ${\cal X}^{\text{cqq}}_{\text{ML}}$
may be attributed to the fact that the ML method returns a lower-rank density matrix with weaker entanglement between the system $AA'BB'$ and the environment $E$.

The consistency of KF and ML results was verified by calculating the Mahalanobis distance \cite{AudenaertScheelNJP2009}
between the density matrices produced by both the methods with the KF covariance matrix used as
a metric. The obtained distance $16.8$ is below the value $17.1$ corresponding to a $95\%$ confidence interval.
The KF method allows one to check for the presence of systematic errors: since the mean of the {\em a posteriori} distribution is not forced to be positive definite, its Mahalanobis distance from the mean of the distribution with imposed positivity constraints
is an indicator of possible systematic errors in the measurement process \cite{AudenaertScheelNJP2009}.
For our data this distance is $17.7$, implying that systematic errors are not significant.

In order to extract a secure key from the four photon state we selected randomly one event from each interval when both the qubits $A$ and $B$ were measured in the $\sigma_y$ bases obtaining $N=3716$ raw key bits. We simulated a binary interactive error-correction procedure \cite{bennett1992} exchanging $990$ parity bits, which corrected all errors, and performed privacy amplification using two-universal hashing functions. Using the KF estimate of Eve's knowledge in the asymptotic limit given by $\chi_E$, conservatively enhanced by five standard deviations, and adding a security margin \cite{Assche2006} to guarantee that the probability of Eve learning at least one bit of the key is below $10^{-6}$, yields $2164$ bits of a secure key.

\begin{figure*}
\includegraphics[width=15cm]{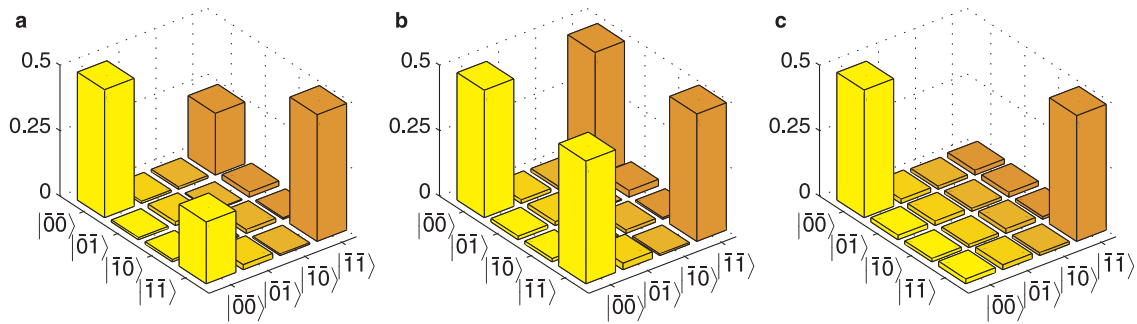}
\caption{Two-qubit $AB$ state. (a) Absolute values of the elements of the reduced density matrix obtained by tracing out qubits $A'$ and $B'$. (b,c) Absolute values of the elements of density matrices conditioned upon finding qubits $A'$ and $B'$ in identical (b) or orthogonal
(c) states when measured in the same basis.}
\label{Fig:twoqubit}
\end{figure*}
The subsystems $A'$ and $B'$ play the role of a shield protecting the private key contained in subsystems $A$ and $B$ from an eavesdropping attempt. Given $\IdealState$, tracing out $A'$ and $B'$ reduces the qubits $A$ and $B$ to a mixed state $\varrho_{AB} =  \frac{1}{4} \proj[AB]{\phi_-} + \frac{3}{4} \proj[AB]{\psi_+}$. The corresponding experimental state, shown in Fig.~\ref{Fig:twoqubit}(a), has ${\cal X}^{\text{cqq}}_{\text{KF}} = -0.009(4)$, which demonstrates that the shield is critical to ensure security. The shield qubits can be used to implement a simple entanglement distillation protocol for $\IdealState$: if $A'$ and $B'$ are projected in the same basis, identical outcomes collapse the state of qubits $A$ and $B$ to a maximally entangled state $\ket[AB]{\psi_+}$, while opposite results produce a separable state $\frac{1}{2} \bigl( \proj[AB]{\phi_-} + \proj[AB]{\psi_+} \bigr) = \frac{1}{2}(\proj[AB]{\bar{0}\bar{0}}+\proj[AB]{\bar{1}\bar{1}})$ useless for
key generation. Fig.~\ref{Fig:twoqubit}(b,c) depict experimental conditional density matrices reconstructed for these two cases using the KF method. The key rate is positive only for identical outcomes and equals $0.693(9)$, which multiplied by the relative frequency of these events 0.511 yields the average value ${\cal X}^{\text{cqq}}_{\text{KF}} = 0.354(5)$, falling significantly behind the result reported in Eq.~(\ref{Eq:ExpSeparation}). Using the resulting subset of qubit pairs to generate a key under the same security assumptions as before yields below 650 bits after error correction and privacy amplification of 1859 raw bits obtained from intervals when the qubits $A$ and $B$ were measured in the same bases. Note that the 50\% reduction in the raw key length compared to the four-photon key extraction corresponds exactly to the success rate of the distillation protocol which halves the raw bit rate if only compatible measurements yielding perfectly correlated outcomes are applied.

In conclusion, we demonstrated experimentally a fundamental feature of private states, namely the separation between distillable entanglement and the secret key contents, using a noisy entangled state of photon quadruplets. The results confirmed the sub-optimality of distillation-based strategies to extract private correlations.
This highlights the complex nature of mixed entanglement in higher dimensions similarly to that exhibited in multiparty scenarios \cite{AmselemBourennaneNPH2009} 
and paves the way to develop QKD protocols that make optimal use of realistic imperfect resources.

We wish to acknowledge insightful discussions with Koenraad Audenaert and Jan Tuziemski. This work was supported by FP7 FET project CORNER, the Foundation for Polish Science TEAM project, and Polish Ministry for Scientific Research project.

\end{document}